\documentclass[twocolumn,preprintnumbers, amsmath,reprint,titlepage,superscriptaddress,10pt,aps,prl]{revtex4-1}

\usepackage{graphicx}
\usepackage{color}
\usepackage{hyperref}
\usepackage{subfigure}
\usepackage{graphicx}
\usepackage{type1cm}
\usepackage{eso-pic}
\usepackage{color}
\usepackage{amsmath}
\usepackage{amssymb}
\usepackage[left]{lineno}
\usepackage{blindtext}
\usepackage{float}

\def\blue#1{\textcolor{black}{#1}}
\def\blue2#1{\textcolor{black}{#1}}

\begin{document}

\title{Experimental triple-slit   interference in a \blue2{strongly-driven  V-type artificial atom}}


\affiliation{SUPA, Institute for Photonics and Quantum Sciences, School of Engineering and Physical Sciences, Heriot-Watt University, Edinburgh EH14 4AS, United Kingdom}
\affiliation{Center for Opto-Electronic Convergence Systems, Korea Institute of Science and Technology, Seoul, Korea}

\author{Adetunmise C. Dada}
 \email{A.C.Dada@Bristol.ac.uk}
\altaffiliation{Current address: Centre for Quantum Photonics, H.H. Wills Physics Laboratory and Department of Electrical and Electronic Engineering, University of Bristol, Bristol, BS8 1TL, United Kingdom}
\affiliation{SUPA, Institute for Photonics and Quantum Sciences, School of Engineering and Physical Sciences, Heriot-Watt University, Edinburgh EH14 4AS, United Kingdom}
\author{Ted~S.~Santana}
\author{Antonios Koutroumanis} 
\affiliation{SUPA, Institute for Photonics and Quantum Sciences, School of Engineering and Physical Sciences, Heriot-Watt University, Edinburgh EH14 4AS, United Kingdom}
\author{Yong~Ma}
 \altaffiliation{Current address: Chongqing Institute of Green and Intelligent Technology, Chinese Academy of Sciences, Chongqing 400714, China}
\affiliation{SUPA, Institute for Photonics and Quantum Sciences, School of Engineering and Physical Sciences, Heriot-Watt University, Edinburgh EH14 4AS, United Kingdom}
\author{Suk-In Park}
\affiliation{Center for Opto-Electronic Convergence Systems, Korea Institute of Science and Technology, Seoul, Korea}
 \author{Jin D. Song} 
\affiliation{Center for Opto-Electronic Convergence Systems, Korea Institute of Science and Technology, Seoul, Korea}
\author{Brian D. Gerardot}
\email{b.d.gerardot@hw.ac.uk}
\affiliation{SUPA, Institute for Photonics and Quantum Sciences, School of Engineering and Physical Sciences, Heriot-Watt University, Edinburgh EH14 4AS, United Kingdom}

\begin{abstract}
Rabi oscillations of a two-level atom appear as a quantum interference effect between the amplitudes associated to atomic superpositions, in analogy with the classic double-slit experiment which manifests a sinusoidal interference pattern.  By extension, through direct detection of time-resolved resonance fluorescence from a quantum-dot neutral exciton driven in the Rabi regime, we experimentally demonstrate triple-slit-type  quantum interference via quantum erasure   in a V-type three-level artificial atom.  This result is of fundamental interest in the experimental studies of the properties of V-type 3-level systems and may pave the way for further insight into their coherence properties as well as  applications for quantum information schemes. It also suggests quantum dots as candidates for multi-path-interference experiments for probing foundational concepts in quantum physics.    
\end{abstract}

\maketitle
\begin{figure}[h]
\includegraphics[width=0.85\linewidth]{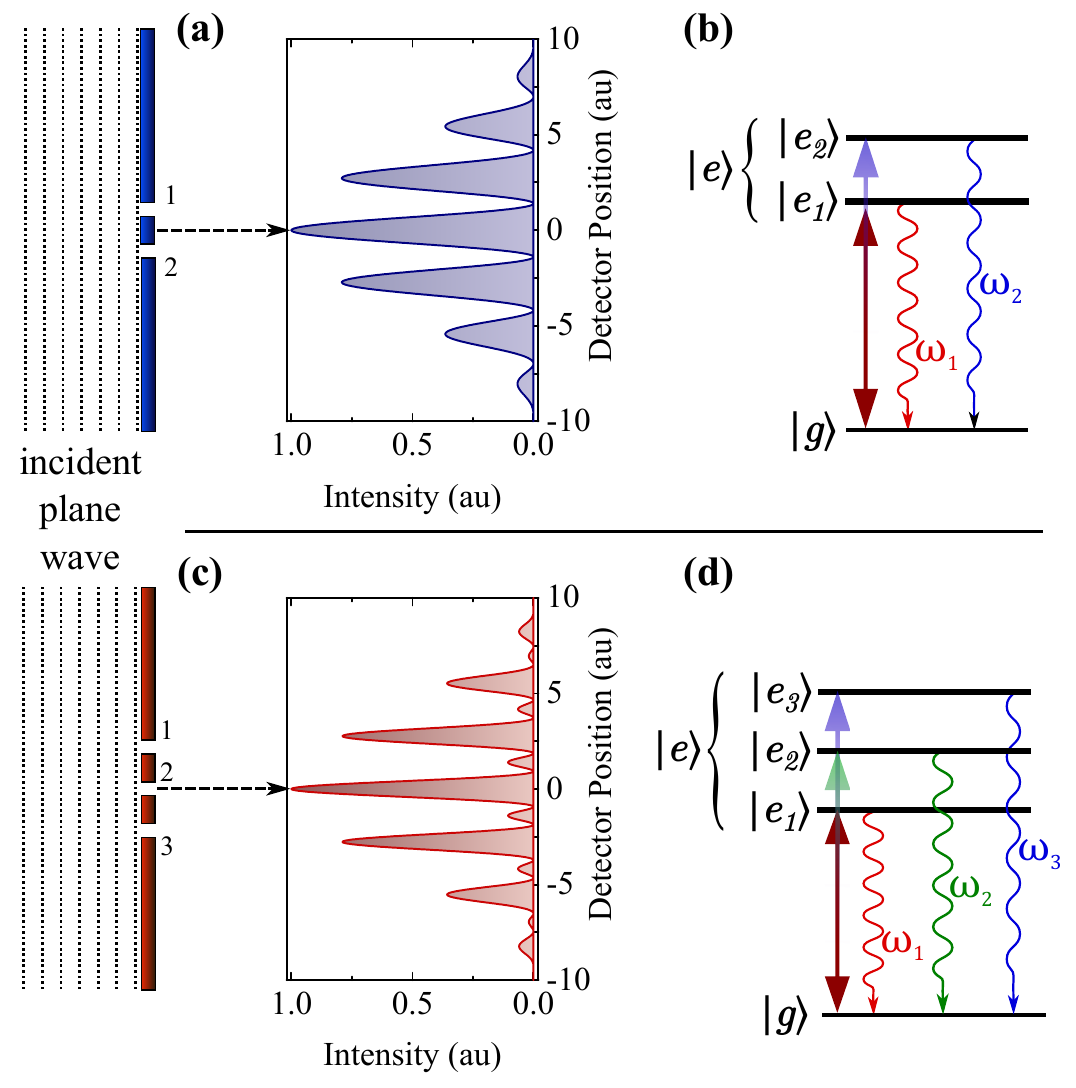}
\caption{{\bf Illustration of double- and triple-slit interference and generic atomic analogues.} 
{\bf (a)} Optical double-slit interference experiment where the  interference pattern has \emph{one} sinusoidal component,  {\bf (b)} Double-slit-type interference setup in an atomic system. The time-domain measurement of fluorescence intensity  will show show oscillations with \emph{one}  beat note {\bf (c)} Optical triple-slit interference experiment  where the  interference pattern has \emph{three} sinusoidal components. {\bf (d)} \blue2{Simplest configuration of an analogue} triple-slit-type interference setup in an atomic system. The time-domain measurement of fluorescence intensity  will show oscillations with \emph{three}  beat notes.
 In (b) and (d), the photon emerges from a superposition of excited states with different energies, where $|e_i\rangle$  ($i=1,2,3$) represent the excited states and $|g\rangle$ the ground state. The energy level structure depicted here is for illustrative purposes only and is not an exact representation of the multilevel configuration in \blue2{our~QD}. }
\label{fig1:illustrslits}
\end{figure}
 
 Quantum interference effects~\cite{PhysRevLett.59.2044, RevModPhys.77.633} form the basis of many photonic quantum information tasks, such as gate operations for  quantum computing~\cite{knill2001scheme,kok2007linear,aaronson2010computational}, quantum state comparison and amplification (see, e.g., \cite{Donaldson:2015gs} and Refs within) and quantum teleportation schemes for the robust transfer of quantum information~\cite{gao2015coherent} essential for quantum networks~\cite{kimble2008quantum}.   
  Although Young's double slit experiment provides an iconic demonstration of quantum interference with photons~\cite{grangier1986}, two-slit-type  interference has also been observed with other (quasi-) particles ~\cite{carnal1991young,schuster1997phase,shin2004atom,zia2007surface}.  Moreover, \emph{\blue2{triple}-slit}-type interference has recently attracted great interest as a tool for probing fundamental questions in  physics, e.g.,  in studies of quantum non-locality~\cite{Niestegge:2013cv},  on the existence of multi-order interference in quantum mechanics~\cite{Sinha:2010kna},  and on the effect of Feynman nonclassical paths in quantum interference experiments~\cite{Sawant:2014eg,daPaz:2016ii}. 
However, \blue2{triple}-slit-type interference has only been demonstrated in optical systems which do not offer the required regime for resolving the latter question~\cite{Sawant:2014eg}. This makes it necessary and interesting to investigate the possibility of multi-slit-type interference in other  physical systems.
 
The phenomenon of \emph{quantum beats}, i.e., a superposed oscillatory behaviour in the light intensity emitted by an atomic system, is a key manifestation of quantum interference and coherence of the underlying states.   A commonly observed quantum-interference phenomenon is Rabi oscillations---understood as a `quantum beat' between two dressed states separated by the Rabi splitting (RS) energy in a driven two-level  (artificial) atom~\cite{Haroche777}. It is a signature of quantum coherence fundamental to the manipulation of atomic qubits in quantum information processing, and has  been observed in quantum dots (QDs)---artificial atoms that can mimic the behaviour of a two-level atom~\cite{Stievater:2001kr,Kamada:2001ila,Htoon:2002ee,Melet:2008bg,Schaibley:2013bt,Dada:2016co}. 
 Another quantum-beat phenomenon occurs  due to interference between the excited states of a V-type atomic system~\cite{macek1969interference,hegerfeldt1994quantum} and has also been observed in the transient decay of QD emission from a suddenly excited neutral exciton state $X^{0}$ (an effective V-system due to fine-structure splitting (FSS)~\cite{bayer2002fine,hogele2004voltage}) using pump-probe setups and quasi-resonant~\cite{Langer:1990cf,Flissikowski:2001hx,PhysRevB.84.195401,Siarry:2015ila} as well as resonant excitation~\cite{Gao:2013df}. 
     FSS quantum beats have been proposed as a means to measure topological phases in Rabi oscillations~\cite{Agarwal:1988be}, also making it interesting to study the FSS beats in the Rabi regime. 
Despite much related theoretical work 
~\cite{PhysRevA.45.505,Zhou:1996ga,Carreno:2003dn,Xue:2016dhba} 
and the availability of various platforms in which V-type systems could be realised, the observation of Rabi oscillations in resonance fluorescence (RF) from a coherently driven  {$V$-type system} remains a fundamentally interesting open question.  In particular, quantum interference involving \emph{both} RS and FSS has not been demonstrated experimentally.  

\begin{figure*}[t!]
\includegraphics[width=0.85\linewidth]{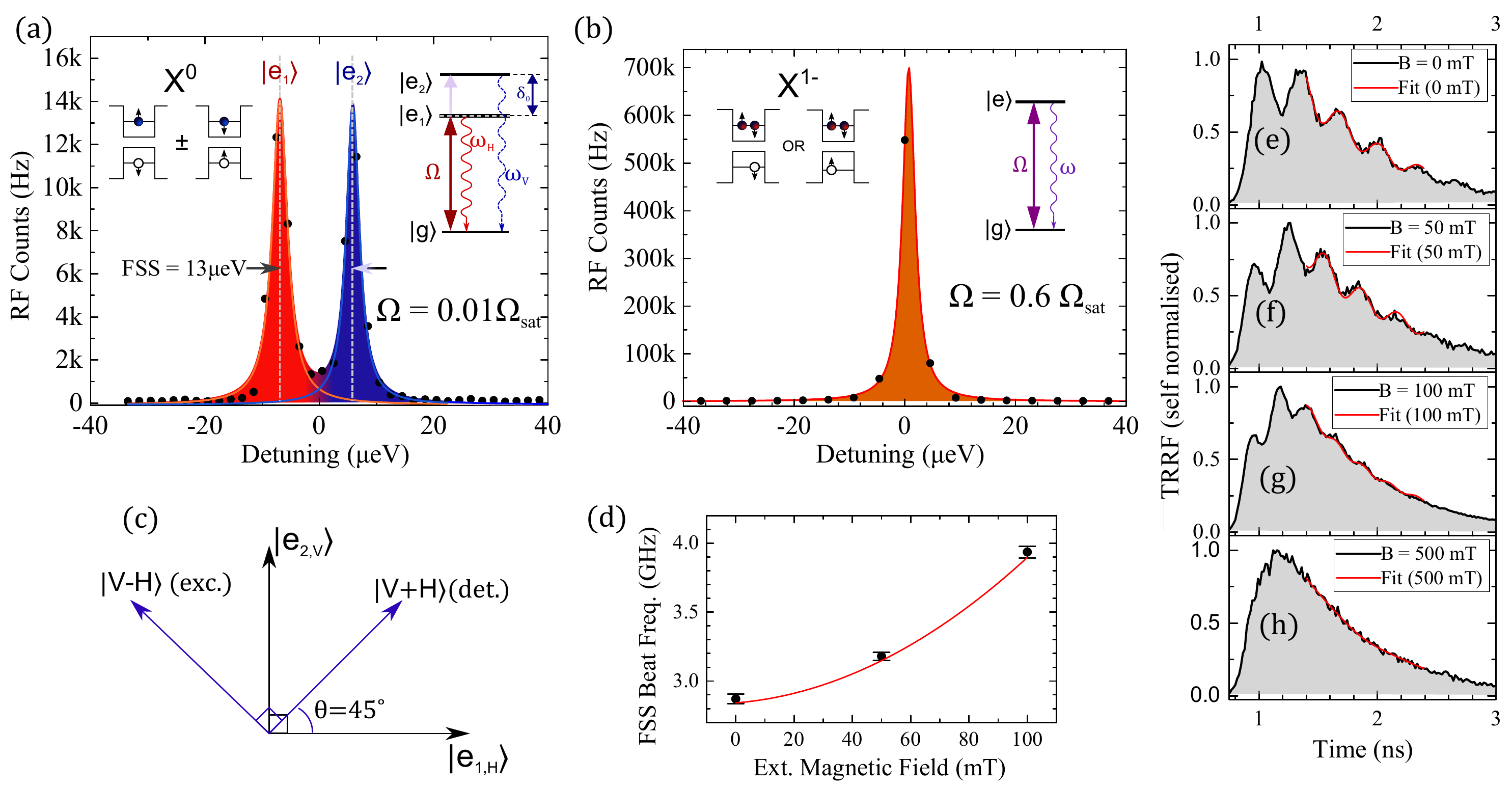}
\caption{{\bf Double-slit-type excitonic interference: FSS quantum beats. (a)}  Resonance-fluorescence (RF) detuning spectra in CW mode for  $X^{0}$,   and {\bf (b)} $X^{1-}$   for comparison. The filled black circles and lines are experimental data and Lorentzian fits, respectively. 
The insets show the corresponding electron and hole spin configurations which have degenerate energies at zero external magnetic field for  $X^{1-}$, effectively making it a two-level system. For  $X^{0}$ 
the electron-hole exchange interaction causes a fine-structure splitting (FSS) of \blue2{$\sim13\mu$eV} at $B=0$ in the QD under study.   {\bf (c) } Quantum erasure scheme.  Excitation and detection polarisation configuration leading to quantum erasure of which-path information originally encoded in the polarization of photons emerging from $|e_1\rangle$, $|e_2\rangle$.  %
{\bf (d)} Quantum-beat frequency versus external magnetic field.  {\bf (e$-$h)}  {$\bf X^0$ 
quantum beats}  under $\pi$-pulse excitation with 100-ps and $\sim5\mu$eV temporal and spectral width, observed at various FSS values manipulated by an external magnetic field $B$. The fits to the data are of the form $I(t)\propto e^{t/T_1}[A+B \cos(\delta_0 t)]$ \cite{Flissikowski:2001hx}. \blue2{Time}-resolved measurements were performed with a $\sim$$150$-ps-resolution detection setup.}
\label{fig2:RFspectra}
\end{figure*}

\begin{figure*}
\includegraphics[width=0.925\linewidth]{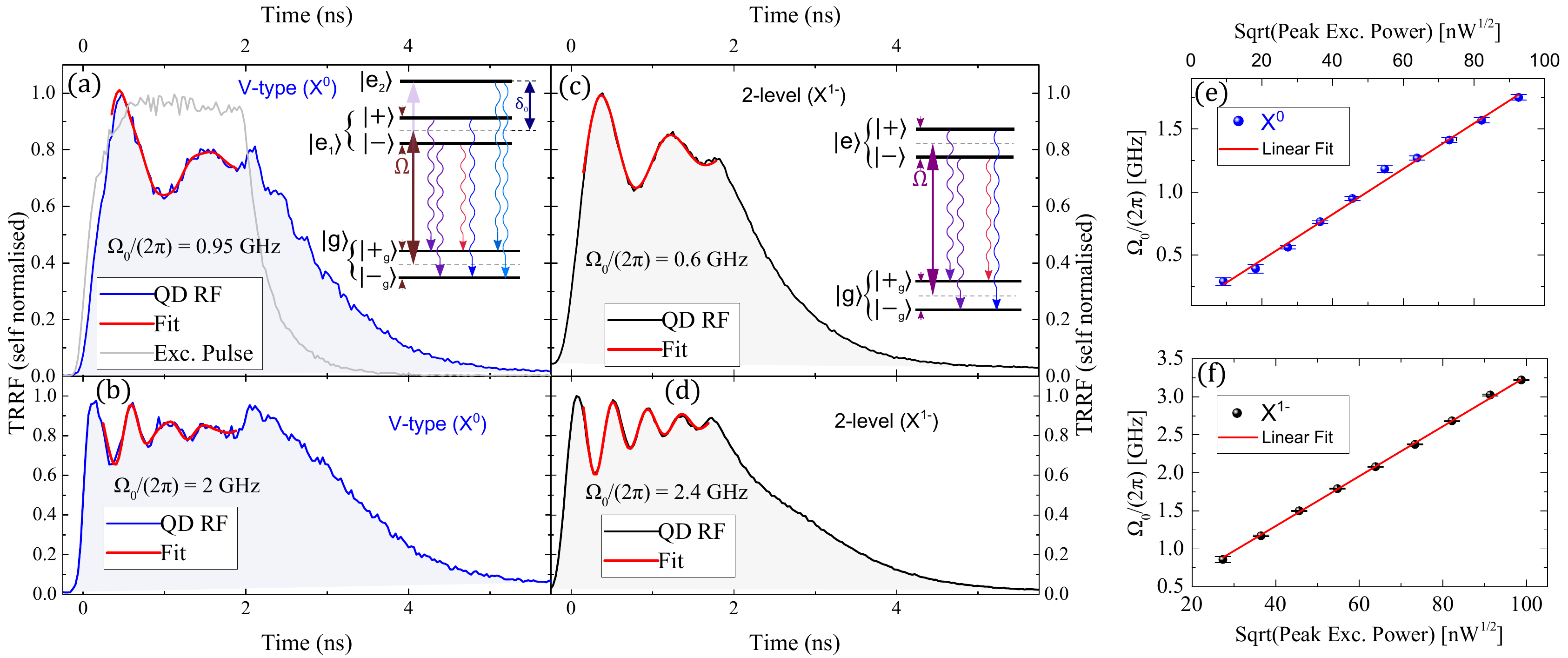}
\caption{{\bf Triple-slit-type excitonic quantum interference:  FSS and RS combined quantum beats}. \blue2{Time}-resolved resonance fluorescence (TRRF) for {\bf (a,b)} driven  $X^0$ exciton (V-system) showing triple-slit-type interference, combining beats due to both Rabi splitting (RS) and fine-structure splitting (FSS), and {\bf (c,d)} the case of a driven  $X^{1-}$ (2-level system) showing double-slit-type interference based only on RS quantum beats, shown for comparison. 
The solid red lines are fits of sinusoidal oscillations to the data having one beat component in (c,d) and three beat components with respective frequencies $|\Omega\pm\delta_0/2|$ and $\Omega$ in (a,b).  The insets illustrate the effective three- and two-slit energy level configurations\blue2{, with the   dressed states defined as  $|\pm\rangle \equiv |g,n+1\rangle\pm |e_{(1)},n\rangle,~|\pm_g\rangle \equiv |g,n\rangle\pm |e_{(1)},n-1\rangle $, where $n$ is the photon number.} {\bf (e,f)}  show Rabi frequencies $\Omega$ extracted from corresponding fits to the data as a function of excitation power for  $X^{0}$ (with FSS \blue2{$\hbar\delta_0=13 \mu$eV}) and $X^{1-}$, respectively. The presence of multiple sinusoidal components evidenced by clear deviation from single sinusoidal oscillations in the time-resolved fluorescence for $X^{0}$ indicates genuine multi-slit type interference.  } 
\label{fig3:qbeats2ns}
\end{figure*}

\begin{figure}
\includegraphics[width=0.9\linewidth]{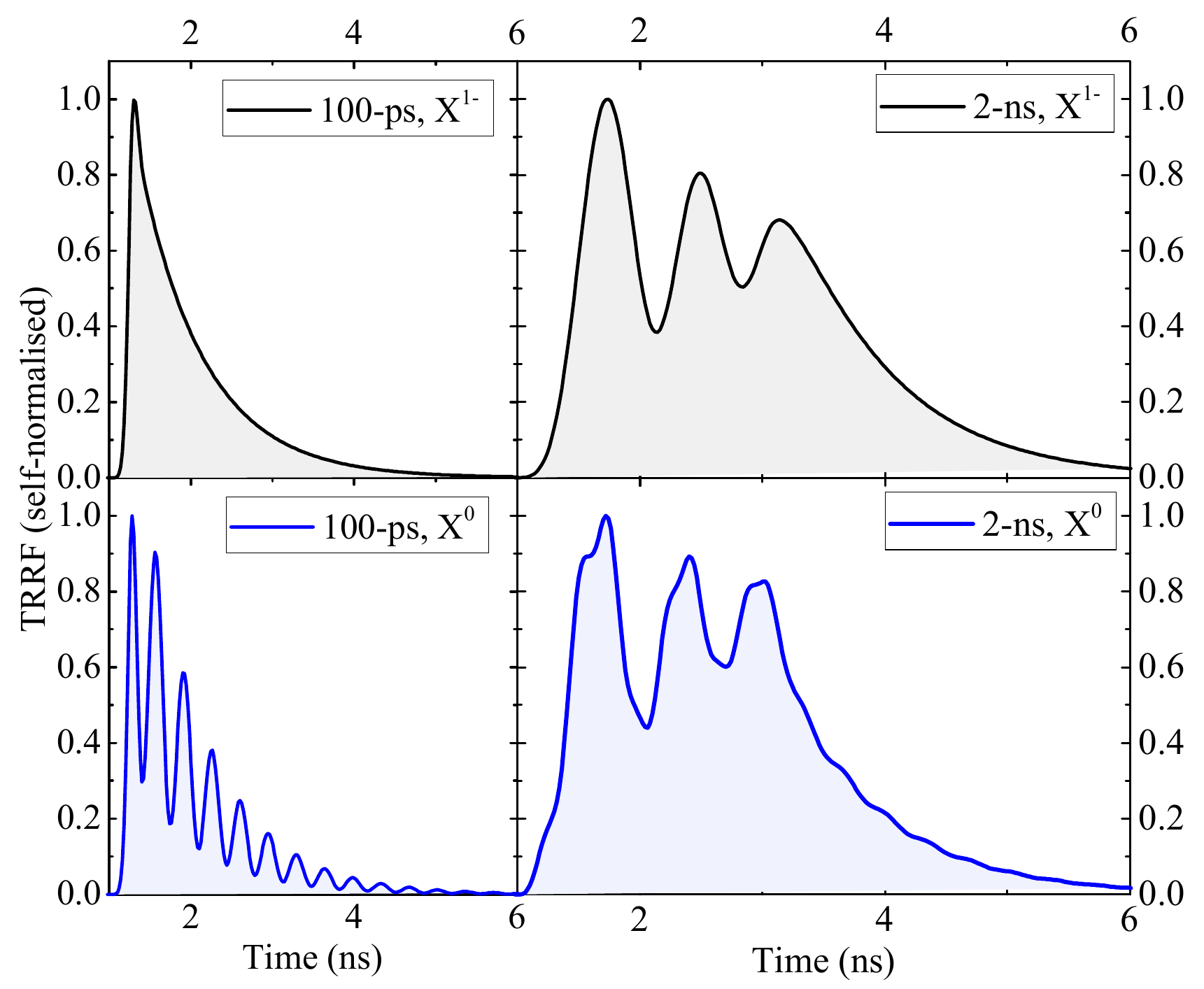}
\caption{{\bf Numerical simulation of TRRF showing RS and FSS quantum beats as well as their combination}.  Simulation of TRRF using the master-equation method with {\bf (a)} 100-ps $\pi$ pulse, $X^{1-}$ {\bf (b)} 100-ps $\pi$ pulse, $X^{0}$  {\bf (c)} 2-ns pulse, $X^{1-}$  {\bf (d)} 2-ns pulse, $X^{0}$. Here \blue2{$\Omega/(2\pi)=1.3$} GHz. }
\label{fig4:qbeats100ps}
\end{figure}

 Here we address these via direct detection of oscillations in time-resolved resonance fluorescence (TRRF) from a V-type system ($X^{0}$ in an InGaAs QD) driven in the Rabi regime, and demonstrate an effective multi-slit interference phenomenon via the combination of RS and FSS. 
  Fig.~\ref{fig1:illustrslits} illustrates the basic idea of double and triple slit interference  and generic atomic analogues. The key signature of multi-slit interference is the presence of  more than one sinusoidal component or beat note in the generated interference pattern. We create an analogous effect in a QD excitonic system with energy configuration effectively of a form similar to the structure in Figs.~\ref{fig1:illustrslits} (b,d). 

QDs are essentially lower band-gap quasi-zero-dimensional structures embedded in a higher-band-gap material.   
Owing to attractive Coulomb  interaction, a trapped electron and the hole form an exciton. Due to the electron-hole exchange interaction, the neutral exciton is energetically split into a doublet (the FSS) with orthogonally linearly-polarized selection rules~\cite{bayer2002fine}. By adding a second electron to form a three-particle trion, the electrons form a spin singlet and the exchange interaction energy (and FSS) vanishes~\cite{hogele2004voltage}.  
For our experiment, we use a self-assembled InGaAs QD embedded in a GaAs Schottky diode for deterministic charge-state control. The QD and experimental setup are described in further detail in Ref.~\cite{Dada:2016co}.  We  focus here on the $X^0$ and show $X^{1-}$ as an effective  two-level-system (TLS) comparison. Figs~\ref{fig2:RFspectra} (a,b) show the resonance fluorescence detuning spectra obtained under CW excitation by tuning the QD through resonance with a narrow-linewidth excitation laser for $X^{0}$ and $X^{1-}$. Being a V-system, the $X^{0}$ exhibits two peaks with a FSS (\blue2{$\sim13\hspace{1pt}\mu$eV}). The single ground state $|g\rangle$ is coupled to two excited states $|e_1\rangle$ and $|e_2\rangle$. \blue2{ Note that the QD energy configuration shown in the insert of Fig.~\ref{fig2:RFspectra} (a) is the same as that illustrated in Fig.~\ref{fig1:illustrslits} (c)}. Due to orthogonal dipole moments of the two transitions, $|e_1\rangle$ and $|e_2\rangle$  decay into orthogonal polarisation modes making the decay paths distinguishable~\cite{Flissikowski:2001hx,bayer2002fine,hogele2004voltage}. This results in a total RF intensity which is proportional to the total excited state population since 
\begin{align}
\langle\hat{n} \rangle =  \langle(\hat{\sigma}^\dag_1\hat{H}+\hat{\sigma}^\dag_2\hat{V})(\hat{\sigma}_1\hat{H}+\hat{\sigma}_2\hat{V})\rangle
= \rho_{11} + \rho_{22},
\end{align}
  which shows no FSS quantum beat~\cite{Flissikowski:2001hx}. Here, the atomic raising and lowering operators are defined as $\hat{\sigma}^\dag_i = |e_i\rangle\langle g|$ and $\hat{\sigma}_i = |g\rangle\langle e_i|$ respectively, while $\hat{\rho}_{ij}$ are elements of the density matrix.

To obtain the  FSS quantum beat, it is necessary to perform  \emph{quantum erasure} of the `which-path' information encoded in the polarisation of emitted photons. For simplicity, we define the neutral exciton states $|e_1\rangle$ and $|e_2\rangle$   as having the polarizations  $H$ and $V$, respectively. The spectral width of the excitation pulses (FWHM$\sim 5\hspace{1pt}\mu$eV) can lead to partial excitation of the \blue2{$|e_2\rangle$}  state even when on resonance with \blue2{$|e_1\rangle$}.  As depicted in Fig.~\ref{fig2:RFspectra}(c), we achieve quantum erasure by detecting the RF through a polarizer which (probabilistically) projects the orthogonal states to the same polarisation, erasing the which-path polarisation information for the filtered photons as also done in Refs~\cite{Langer:1990cf,Flissikowski:2001hx,PhysRevB.84.195401,Siarry:2015ila,Gao:2013df}. The same \blue2{polarizer  suppresses} the scattered excitation laser light in our RF setup. 
The projected RF intensity is then proportional to 
\begin{align}
\langle\hat{n}\rangle_{\rm det} \propto \rho_{11} + \rho_{22}+ {\rho}_{12}+ {\rho}_{21},
\end{align}
which now includes \emph{coherence} terms 
representing interferences between the two excited-state populations.  
 For Fig.~\ref{fig2:RFspectra}(d-h),  we excite the $X^0$ QD transition with  100-ps $\pi$ pulses on resonance with the $|g\rangle \leftrightarrow |e_1\rangle$ transition and measure the TRRF. We observe that the beat frequency corresponds to the FSS energy and that it changes with the FSS as manipulated \blue2{with an} external magnetic field $B$~\cite{PhysRevB.84.195401,Siarry:2015ila} [Fig.~\ref{fig2:RFspectra}(d)], confirming the observed oscillations as quantum beats due to FSS in \blue2{$X^0$}.    Fits to the transients as shown in \ref{fig2:RFspectra}(e-h), corroborated by Fourier transformations, show that the oscillations have a single sinusoidal component, consistent with double-slit-type interference. The interference visibility reduces with increasing $B$ due to a combination of two effects: (1) the increased oscillation frequency is less resolvable with the finite resolution of the detection setup ($\sim$150\hspace{1pt}ps FWHM); (2) the superposition created is less equal (other work in which nearer equal superposition is created using a two-color scheme shows higher visibility~\cite{Gao:2013df}). In our experiment, the preparation of the superposition is accomplished by simultaneous excitation of the excited states since the 100-ps pulses are short compared to the reciprocal of the FSS frequency.
 We note however that an initial superposition state is not strictly necessary for the occurrence of FSS quantum beats~\cite{hegerfeldt1994quantum}. 

The demonstration of \blue2{multi-slit-type interference is shown in Fig.~\ref{fig3:qbeats2ns}, where we clearly see a combined effect of RS \blue2{($\hbar\Omega$)}  and FSS \blue2{($\hbar\delta_0$)}. The three dominant beat frequencies in the time domain due to this combination of splittings are $|\Omega\pm\delta_0/2|$ and $\Omega$.} A fit with 3 sinusoidal components \blue2{at  these beat frequencies} consistently gives a good fit to the TRRF data at various excitation powers (\blue2{corresponding to various Rabi frequencies $\Omega$).  Of these,} we plot two examples in Fig.~\ref{fig3:qbeats2ns}(a) and (b) \blue2{and we illustrate the combination of $\Omega$ and $\delta_0$ in the inset of Fig.~\ref{fig3:qbeats2ns}(a) to highlight the source of the beat frequencies. We note that although the QD energy configurations in the insets of Fig.~\ref{fig3:qbeats2ns}(a,b)  differ  slightly from Fig.~\ref{fig1:illustrslits}(c,d), respectively,  due to the existence of two ground state levels $|+\rangle_g$, the fact that their splitting is the same as that between two of the excited-state levels makes the interference effect equivalent to the basic double- and triple-slit cases illustrated in Fig.~\ref{fig1:illustrslits}(c,d)}. For this demonstration we used 2-ns pulses for excitation to accommodate slower  oscillation frequencies. Similar measurements on the $X^{1-}$ of the same dot allows us to make direct comparison with Rabi oscillation in a \blue2{two-level system (TLS)} in Fig.~\ref{fig3:qbeats2ns}(c) and (d).   Assuming constant FSS at the various excitation powers, we obtain up to 3 sinusoidal components of the oscillations in TRRF from which we extract the Rabi frequencies plotted in Fig.~\ref{fig3:qbeats2ns}(e). We compare the   RS beats for a TLS in (f). The presence of \blue2{at least} 3 beat components is in agreement with previous theoretical studies of the \blue2{time-domain} dynamics~\cite{Xue:2016dhba} and power spectrum for a driven $V$-system which is predicted to show \blue2{more than 3 peaks  (i.e. 5 or 7 peaks)~\cite{PhysRevA.45.505,Zhou:1996ga,Carreno:2003dn}. These consist of the central peak, i.e., carrier frequency, and pairs of side peaks which cause extra time-domain beat components.}

We note that three `slits' are created by Rabi splitting of $|e_1\rangle$, one of the $X^0$  doublet lines, with which the laser is on resonance,  into two dressed states $|+\rangle$, $|-\rangle$ [see inserts in Fig.~\ref{fig3:qbeats2ns}(a) and (c)]. In this case, polarisation-encoded which-path information in the emitted light  is partial as it can only distinguish between $|e_2\rangle$ and \{$|+\rangle$ or $|-\rangle$\}, not between $|+\rangle$ and $|-\rangle$.  Quantum erasure of this information is carried out by the polarisation filtering described above \blue2{[see Fig.\ref{fig2:RFspectra}(c)].  A  feature of this three-slit-type interference is the tunability of the interslit distances} via the excitation power and   \blue2{ $B$, respectively}.


We model this effect using the master-equation method~\cite{loudon2000quantum,scully1997quantum} (as in our previous work~\cite{Dada:2016co}), and obtain the TRRF intensity through the polarizer.  Fig.~\ref{fig4:qbeats100ps} shows the simulation results obtained by modelling excitation of  the $|g\rangle\leftrightarrow|e_1\rangle$ transition using ~either 100-ps $\pi$ pulses (a-b) to demonstrate the FSS beats or 2-ns pulses (c-d) to demonstrate the combination of RS and FSS beats.   The simulation results show clear qualitative agreement with the measured experimental data (Fig.~\ref{fig3:qbeats2ns}). In Fig.~\ref{fig4:qbeats100ps} (d) we see a modified Rabi oscillation pattern, containing extra beat components due to the combination of RS and FSS.  From the simulations, we also verify that both the FSS and RS beats originate in the evolution of the exciton itself.  
The overall deviation from a single sinusoidal oscillation is a clear signature of multilevel quantum coherence and at least 3-slit-type quantum interference.  
On its own, the FSS beat demonstrates superposition between the two excited doublet states. In turn, the Rabi driving field creates evolution of superpositions between the ground (no exciton)  and excited state as a function of time. The combination of these two create a coherent superposition of multiple quantum states evolving as a function of time. 

These results are of fundamental interest because \blue2{TRRF from strongly driven V-type atom has not previously been demonstrated. Moreover,  we show that it can effect genuine multi-slit-type quantum interference. We note briefly that multi}-slit experiments proposed to quantify contributions from nonclassical paths in quantum interference  need to operate in a regime where the deviation from a naive application of the superposition principle is measurable, and this is known to be difficult to achieve in \blue2{optical setups. The contribution of nonclassical paths to the interference pattern} is quantified using a parameter $\kappa$ which depends on the slit thickness, inter-slit distance and photon wavelength, amongst other experiment parameters. For optical setups, it has been shown that $\kappa$ increases with photon wavelength when other experiment parameters are fixed~\cite{Sawant:2014eg}. \blue2{In our analogous setup, we expect that the linewidths of the transitions ($\sim 1 \mu$eV), splitting energies ($\hbar \Omega$, $\hbar \delta_0$$\sim$$ 13 \mu$eV), and photon energies  (e.g., $\sim 1.3 $eV) would represent the slit widths, inter-slit distances and photon wavelengths in optical setups, respectively.}
This  suggests that large values of $\kappa$ might be obtained from an analogous multi-slit interference such as demonstrated here since the \blue2{photon energy} relative to \blue2{our} `slit thickness' and `inter-slit distance' is orders of magnitude larger than  in proposed optical setups. 
In future work, it would be interesting to investigate in detail the effective values of the $\kappa$ parameter in such an excitonic multi-path setup and the resulting feasibility of such experiments.

In conclusion, we have demonstrated an analogue of optical \blue2{triple-slit quantum interference via direct detection of time-resolved resonance fluorescence from a strongly driven V-type artificial atom, which has not previously been  investigated in any atomic system}.  
This could potentially lead to new fundamental investigations in quantum mechanics. Further, understanding the properties of the V-type atomic systems is likely to be important for applications in future quantum technologies, and here we have demonstrated a key feature of such a \blue2{system when} driven in the Rabi regime.


{\bf Funding and Acknowledgements.} The Authors would like acknowledge the financial support for this work from a Royal Society University Research Fellowship, the EPSRC (grant numbers EP/I023186/1, EP/K015338/1, and EP/G03673X/1) and an ERC Starting Grant (number 307392). KIST authors acknowledge support from KIST's flag-ship program. ACD acknowledges helpful  discussions and support from Ruth Oulton.



%



\end{document}